\documentclass[prd,onecolumn,aps,nofootinbib]{revtex4-2}
\usepackage{graphicx}
\usepackage{epstopdf}
\usepackage{hyperref}
\usepackage{latexsym}
\usepackage{amsmath}
\usepackage{amssymb}

\usepackage[utf8]{inputenc}
\hypersetup{
    colorlinks=true,
    linkcolor=red,
    citecolor=blue,
}
\usepackage{color}
\usepackage[T1]{fontenc}
\usepackage{txfonts}
\usepackage{supertabular}
\usepackage{setspace}
\usepackage{multirow}
\usepackage{makecell}
\usepackage{mathrsfs}
\usepackage{subfig}

 \newcommand{\bq}{\begin{equation}}
 \newcommand{\eq}{\end{equation}}
 \newcommand{\bqn}{\begin{eqnarray}}
 \newcommand{\eqn}{\end{eqnarray}}

\makeatletter

\newcommand{\Rmnum}[1]{\expandafter\@slowromancap\romannumeral #1@}

\makeatother

\begin{document}

\title{Influence of dark matter equation of state on the axial gravitational ringing of supermassive black holes}

\author{Yuqian Zhao}
\email[Yuqian Zhao: ]{yqzhao@mail.bnu.edu.cn}
\affiliation{Institute for Frontiers in Astronomy and Astrophysics, Beijing Normal University, Beijing 102206, China}
\affiliation{Department of Astronomy, Beijing Normal University, Beijing 100875, China}
\author{Bing Sun}
\affiliation{Department of Basic Courses, Beijing University of Agriculture, Beijing 102206, China}
\affiliation{CAS Key Laboratory of Theoretical Physics, Institute of Theoretical Physics, Chinese Academy of Sciences, Beijing 100190, China}
\author{Zhoujian Cao}
\email[Zhoujian Cao (corresponding author): ]{zjcao@amt.ac.cn }
\affiliation{Institute for Frontiers in Astronomy and Astrophysics, Beijing Normal University, Beijing 102206, China}
\affiliation{Department of Astronomy, Beijing Normal University, Beijing 100875, China}
\affiliation{School of Fundamental Physics and Mathematical Sciences, Hangzhou Institute for Advanced Study, UCAS, Hangzhou 310024, China}

\author{Kai Lin}
\affiliation{
Hubei Subsurface Multi-scale Imaging Key Laboratory, School of Geophysics and Geomatics, China University of Geosciences, Wuhan 430074, Hubei, China
}

\author{Wei-Liang Qian}
\email[Wei-Liang Qian: ]{wlqian@usp.br}
\affiliation{Escola de Engenharia de Lorena, Universidade de S\~ao Paulo, 12602-810, Lorena, SP, Brazil}
\affiliation{Faculdade de Engenharia de Guaratinguet\'a, Universidade Estadual Paulista, 12516-410, Guaratinguet\'a, SP, Brazil}
\affiliation{Center for Gravitation and Cosmology, College of Physical Science and Technology, Yangzhou University, Yangzhou 225009, China}

\begin{abstract}

In this work, we explore the effects of surrounding dark matter featuring different equations of state on the axial gravitational quasinormal modes of supermassive black holes situated at the center of galaxies. 
Our attention primarily rests on dark matter exhibiting a spike structure, originating from relativistic Bondi accretion through an adiabatic process, which diminishes at a certain distance from the black hole. 
We analyze how varying the equation of state of the dark matter influences the properties of the spacetime in the black hole's vicinity. 
Our findings reveal that different states of dark matter spikes correspondingly affect the black hole's quasinormal modes. 
In particular, we identify deviations in both the ringing frequency and damping time, reaching magnitudes of up to $10^{-3}$ for certain parameter values. 
These variations can potentially be detected by upcoming space-borne detectors. 
Our findings thus indicate the feasibility of discerning and limiting the essential properties of dark matter surrounding supermassive black holes using future gravitational wave detections, particularly in the case of extreme mass ratio inspiral systems.

\end{abstract}

\maketitle

\newpage

\section{Introduction}
 
Black holes (BHs) stand as notable predictions of general relativity (GR)~\cite{einstein1915feldgleichungen}. 
Over the past century, their existence and distinctive characteristics have been the focus of continuous endeavors undertaken by the scientific community~\cite{Barack_2019}.
Since the inaugural detection of a gravitational wave (GW) event in 2015~\cite{PhysRevLett.116.061102}, the field has made remarkable strides, recording more than ninety BHs binary events~\cite{GWOSC,GWTC1,GWTC2,GWTC2.1,GWTC3}. 
Moreover, the observations of the images of the supermassive black holes (SMBHs) located at the center of M87 and our galaxy have further furnished valuable information on the subject~\cite{1Akiyama_2019,2Akiyama_2019,3Akiyama_2019,4Akiyama_2019,5Akiyama_2019,6Akiyama_2019,7Akiyama_2021,8Akiyama_2021}.

The event horizon is an intriguing feature of BH. 
As a one-way causal boundary, it prevents us from detecting the inside of a BH~\cite{1916skpa.conf..424S}. 
In this regard, the quasi-normal modes (QNMs) of BHs~\cite{chandrasekhar1985mathematical,kokkotas1999quasi, Hans-PeterNollert_1999, Berti_2009, RevModPhys.83.793}, in the form of ringdown GWs, constitute one of the potential means to scrutinize the whereabouts and properties of these compact objects. 
In practice, a BH is likely to emerge in an astronomical environment, such as interstellar dust or halo comprised of ordinary and dark matter (DM)~\cite{PhysRevD.96.083014,Nampalliwar_2021,Xu_2021}. 
It is understood that 90\% of the host galaxies of SMBHs are composed of DM~\cite{jusufi2020shadows}. 
Furthermore, two pioneering works \cite{PhysRevD.89.104059,Barausse_2015} first investigated the environmental effects on GWs and found it possible to explore the characteristics of the environment around the BHs through GWs detection in some scenarios.
Therefore, it is interesting to ask how the DM around the SMBHs influences the properties of the spacetime and the GWs emanating from them.

The numerical results from $N$-body cosmological simulations suggest that the density distribution of DM is peaked near the center of galaxies and decreases as a power of $1/r$ with $r$ the distance from the halo's center.
To be specific~\cite{PhysRevD.106.044027},
\begin{equation}\label{rho}
\rho(r)={\bar \rho}(r / {\bar r})^{-\gamma_0}\left[1+(r / {\bar r})^{\alpha_0}\right]^{(\gamma_0-\beta_0) / \alpha_0}\,,
\end{equation}
where ${\bar r}$ and ${\bar \rho}$ are the scale factors which are determined by the numerical fitting as in~\cite{10.1093/mnras/stz1698}.
The parameters, $\alpha_0$, $\beta_0$ and $\gamma_0$, are tailored to specific models.
For example, for Hernquist profile~\cite{hernquist1990analytical}, one has $(\alpha_0,\beta_0,\gamma_0)=(1,4,1)$, while $(\alpha_0,\beta_0,\gamma_0)=(1,3,1)$ are adopted for Navarro-Frenk-White (NFW) profile~\cite{Navarro_1997}.

With the presence of a BH, the density distribution of the DM is modified.
In the pioneering work~\cite{PhysRevLett.83.1719}, a Newtonian method is employed to calculate the distribution of cold DM near the center of galaxies. 
The BH accretion gives rise to forming a cuspy structure, also referred to in the literature as a ``spike''. 
Subsequently, for a spherically symmetric BH, the density peaks near $r\gtrsim 4 R_s$ with $R_s$ the Schwarzschild radius.
The profile is also featured by a steep cutoff towards the inside at $r=4 R_s$, below which the density of DM vanishes due to annihilation or absorption into the BH. 
When the relativistic modifications are taken into account~\cite{PhysRevD.88.063522}, the main features of the density distribution largely remain unchanged, while the cutoff radius is found to assume a smaller value $r=2R_s$.

In~\cite{PhysRevD.106.044027}, both Hernquist and NFW DM spikes from relativistic Eddington accretion are considered to explore the impacts of the DM spike on the extreme mass-ratio inspirals (EMRIs) GW waveforms. 
The relativistic modifications are found to positively impact the DM detectability for both models. 
On the other hand, the impacts of the cold DM halo on the ringdown waveforms, namely, the QNMs are explored in~\cite{PhysRevD.104.104042, ZHANG2022101078,PhysRevD.104.124082, PhysRevD.105.L061501, KONOPLYA2021136734,Konoplya_2022}. 
Besides, the impacts of the DM spike on the QNMs of scalar perturbations and axial gravitational perturbations can be found in~\cite{Daghigh_2022,PhysRevD.108.024070}. 
Particularly, in~\cite{PhysRevD.108.024070}, the DM spike profile from Eddington accretion in \cite{PhysRevD.106.044027} was considered, and the pressure of the DM was neglected for simplicity. 

The present study is motivated to explore further the impact of DM's equations of state (EOSs) on the resulting QNMs of the underlying SMBHs.
In particular, the DM is modeled by an isentropic fluid, whose EOS is governed by the Bondi's form~\cite{1952MNRAS.112..195B,michel1972accretion}:
\begin{equation}
    p(r)=\alpha \rho_0^\gamma\,,
\end{equation}
where $\alpha$ is a constant, $\gamma\in [1,3]$ represents the adiabatic indices of the EOS, and $\rho_0$ denotes the rest-mass density~\cite{10.1093/mnras/stab161}. 
As a result, the DM possesses a finite pressure $p(r)$. 
In particular, we will elaborate on soft and stiff EOSs with, respectively, $\gamma<\frac{5}{3}$ and $\gamma>\frac{5}{3}$~\cite{10.1093/mnras/stab161}.
Regarding the DM profiles, the Bondi accretion effect~\cite{2022JCAP...08..032F} will be considered instead of that due to the Eddington accretion.
Our analysis will also focus on the detectability of ringdown waveforms.

As discussed in~\cite{1952MNRAS.112..195B,10.1093/mnras/stab161}, the value $\gamma=\frac{5}{3}$ can be roughly viewed as a watershed, the dividing point between the Newtonian and relativistic scenarios. 
Specifically, EOSs with $\gamma\leq\frac{5}{3}$ are referred to as soft ones, for which the sound speed is much less than the speed of light.
As a result, the Newtonian framework suffices for a reasonable description of DM's accretion process. 
On the other hand, for stiff EOSs with $\gamma>\frac{5}{3}$, the Newtonian approach might lead to nonphysical solutions and, therefore, the relativistic modifications become indispensable. 
Two other values of interest are $\gamma=2$ and $\gamma=3$, which correspond to, respectively, the two-body and three-body interacting superfluid DM~\cite{DeLuca_2023}. 
We will further elaborate on the behaviors of QNMs near these particular values.

We argue that the obtained results indicate the feasibility of probing the presence of DM spike structure and extracting the information on the DM EOS through the ringdown waveforms.
In this regard, the detectability of the modification to the QNMs owing to different EOSs is discussed. 
Nonetheless, the signal-to-noise ratio (SNR) of the ground-based GW detectors is not favorable for successful observation of the ringdown signals~\cite{PhysRevLett.129.111102}. 
Such detection is likely feasible for the ongoing space-borne detectors such as LISA, TianQin, Taiji, and DECIGO~\cite{ruan2020taiji, PhysRevD.100.044042, Moore_2015}. 
In particular, a recent study~\cite{PhysRevD.100.044036} indicated that a relative deviation of the order $\sim 10^{-3}$ falls within the sensitivity range. 
This point will also be taken into account in our analysis.

The remainder of the paper is organized as follows.
Based on the relativistic adiabatic process, we first introduce Bondi's EOSs and derive the corresponding density profile of the DM spike.
Subsequently, the metric is obtained and presented in Sec.~\ref{sec:BHinDM}. 
We then derive the master equation of the QNMs for axial gravitational perturbations in Sec.~\ref{sec:QNMeqn}. 
In Sec.~\ref{sec:result}, we solve the QNMs equations numerically for the quasinormal frequencies. 
We will focus on the impact of different EOSs on QNMs, particularly for the specific values of $\gamma$ mentioned above. We discuss the detectability of consequential modifications to the QNMs through the waveforms.
The last section is devoted to further discussions and concluding remarks.
Throughout the paper, we use the geometric unit system so that $c=G=1$, where $c$ is the speed of light and $G$ is the gravitational constant.

\section{Black Holes merged in dark matter spike from Bondi accretion}\label{sec:BHinDM}

This section examines the spherically symmetric BH metric that are surrounded by the DM spike from Bondi Accretion. 
We set out to derive the density profile of the DM spike and then obtain the corresponding modified black hole metric. 

Although the Schwarzschild BH is the most well-known solution for the spherically symmetric static spacetime, it only applies to the scenario of an isolated black hole in vacuum. 
The latter is largely improbable in practice, especially for SMBHs located at the center of galaxies. 
These black holes are typically surrounded by a complex distribution of matter, primarily the DM. 
In general, the metric of a Schwarzschild BH merged in DM can be described by the following form:
\begin{equation}\label{metric}
        ds^2 = -f(r)dt^2 + \frac{dr^2}{g(r)} + r^2 (d \theta^2 + \sin^2 \theta d\varphi^2) \, ,
\end{equation}
and the contribution from the DM can be attributed to the energy-momentum tensor:
    \begin{equation}\label{Tmunu}
        T^\mu_\nu=\mathrm{diag}\left\{-\rho(r),p(r),p(r),p(r)\right\} \,,
    \end{equation}
where $\rho(r)$ is the density distribution of DM and the pressure $p(r)$ depends on the EOS. 

The Einstein or TOV equations are found to be
\begin{equation}\label{TOV1}
    \kappa^2 T_t^t=-8 \pi G \rho=\frac{r g^{\prime}+g-1}{r^2} \,,
\end{equation}
\begin{equation}\label{TOV2}
    \kappa^2 T_r^r=8 \pi G p=\frac{g-1}{r^2}+\frac{g f^{\prime}}{r f} \,,
\end{equation}
and
\begin{equation}\label{TOV3}
    \nabla_\nu T_r^\nu=0
\end{equation}
gives
\begin{equation}\label{TOV4}
\frac{d p}{d r}=-(\rho+p) \Gamma_{01}^0=-\frac{1}{2}(\rho+p) \partial_r \ln f \,.
\end{equation}

We therefore obtain a system of three equations that involve four unknown variables, namely $\left[f(r), g(r), \rho(r), p(r)\right]$.
The remaining degree of freedom demands an additional physical condition, which can be fixed by the DM's EOS.
In~\cite{Daghigh_2022,PhysRevD.108.024070}, the particular choice of DM density profile guarantees that there is no surplus freedom for the EOS. 
Such an approach effectively neglects the pressure of the DM $p(r)$. 
The present study will explicitly consider such a physical ingredient and examine its impact on the resultant black hole QNMs, focusing on any detectable deviations. 
In particular, we will adopt the Bondi's EOS and derive the corresponding density profile below in Sec.~\ref{sec:profile}. The corresponding modifications to the metric will be discussed in Sec.~\ref{sec:gauge}.

\subsection{The DM distribution from Bondi accretion}\label{sec:profile}

Based on ideal relativistic hydrodynamics, the DM is subject to an adiabatic process that yields the relation between the total and rest-mass energy densities in the local rest frame $\rho$ and $\rho_0$~\cite{10.1093/mnras/stab161,baumgarte2010numerical}:
\begin{equation}
    \left(\frac{\partial \rho}{\partial \rho_0}\right)_{\mathrm{ad}}=\frac{\rho+p}{\rho_0}\,.
\end{equation}
By explicitly considering the radial dependence of the profile $\rho_0(r)$, we can rewritten the above relation as:
\begin{equation}\label{ad}
    \frac{d \rho}{d r}=\frac{d \rho_0}{d r} \frac{\rho+p}{\rho_0}\,.
\end{equation}
We now assume the Bondi's EOS given by
\begin{equation}\label{state}
    p=\alpha \rho_0^\gamma \,.
\end{equation}
where $\alpha$ is a constant coefficient and $\gamma$ is the adiabatic indices with the range of $[1,3]$. Therefore, five equations, Eqs.~(\ref{TOV1}-\ref{TOV3}) and~(\ref{ad}-\ref{state}), are accounted for by five unknowns $\left[f(r), g(r), \rho(r), p(r),\rho_0(r) \right]$.

As a reasonable approximation, we proceed by solving the above system of equations in an iterative fashion.
We first consider how the DM distributions are affected by a Schwarzschild BH by taking into account Eqs.~\eqref{TOV3} and~(\ref{ad}-\ref{state}). 
By substituting Eq.~\eqref{state} into Eq.~\eqref{ad}, we have,
\begin{equation}
    \rho=\frac{\alpha  \rho_0^{\gamma }}{\gamma -1}+\rho_0=\left(\frac{p}{\alpha }\right)^{\frac{1}{\gamma}
    }+\frac{p}{\gamma -1}\,.
\end{equation}
Then, considering the Schwarzschild case with $f(r)=g(r)=1-\frac{2M}{r}$, Eq.~\eqref{TOV3} now becomes
\begin{equation}\label{dpdr1}
    \frac{d p}{d r}=-\frac{1}{1-\frac{2 M}{r}}\frac{M}{r^2}\left\{\left[\frac{p(r)}{\alpha }\right]^{\frac{1}{\gamma}}+\frac{\gamma  p(r)}{\gamma -1} \right\}\, .
\end{equation}
Integrating Eq.~\eqref{dpdr1} we have
\begin{equation}
    p(r)=\left(\frac{\gamma -1}{\gamma }\right)^{\frac{\gamma }{\gamma -1}} \alpha ^{-\frac{1}{\gamma -1}}\left(\frac{C_0 \alpha ^{\frac{1}{\gamma}}}{(\gamma -1)\sqrt{1-\frac{2 M}{r}}}-1\right)^{\frac{\gamma }{\gamma -1}}\,.
\end{equation}
By substituting above expression of pressure into Eq.~\eqref{state} and Eq.~\eqref{ad}, respectively, the density profiles $\rho_0(r)$ and $\rho(r)$ are derived which read
\begin{equation}
    \rho_0(r)=\left(\frac{\gamma -1}{\gamma }\right)^{\frac{1}{\gamma -1}} \alpha ^{-\frac{1}{\gamma -1}}\left(\frac{ C_0 \alpha ^{\frac{1}{\gamma} }}{(\gamma -1)\sqrt{1-\frac{2 M}{r}}}-1\right)^{\frac{1}{\gamma -1}}\,,
\end{equation}
\begin{equation}\begin{aligned}
    \rho(r)=\left(\frac{\gamma -1}{\gamma }\right)^{\frac{1}{\gamma -1}} \alpha ^{-\frac{1}{\gamma -1}}\left(\frac{C_0 \alpha ^{\frac{1}{\gamma}}}{(\gamma -1)\sqrt{1-\frac{2 M}{r}}}-1\right)^{\frac{1}{\gamma -1}}+\left(\frac{\gamma -1}{\gamma }\right)^{\frac{\gamma }{\gamma -1}} \frac{\alpha ^{-\frac{1}{\gamma -1}}}{\gamma -1}\left(\frac{C_0 \alpha ^{\frac{1}{\gamma}}}{(\gamma -1)\sqrt{1-\frac{2 M}{r}}}-1\right)^{\frac{\gamma }{\gamma -1}}\,.
\end{aligned}\end{equation}

To determine the constant of integration $C_0$, we examine the asymptotic behavior of the solution at spatial infinity. 
For $r\gg M$, the profile should be governed by a power law form $\rho_0\sim r^{\beta}$, not a constant.
This fixes $C_0$ to be
\begin{equation}
   C_0= (\gamma -1) \alpha ^{-\frac{1}{\gamma }}\,.
\end{equation}
By defining an effective density parameter $\Tilde{\rho}_0$
\begin{equation}
    \Tilde{\rho}_0=\left(\frac{\gamma -1}{\gamma }\right)^{\frac{1}{\gamma -1}} \alpha ^{-\frac{1}{\gamma -1}}\,,
\end{equation}
the density profiles and pressure now read
\begin{equation}\label{rho0}
    \rho_0(r)=\Tilde{\rho}_0\left[\left(1-\frac{2 M}{r}\right)^{-\frac{1}{2}}-1\right]^{\frac{1}{\gamma -1}}\,,
\end{equation}
\begin{equation}\label{pressure}
    p(r)=\Tilde{\rho}_0\frac{\gamma -1}{\gamma}\left[\left(1-\frac{2 M}{r}\right)^{-\frac{1}{2}}-1\right]^{\frac{\gamma}{\gamma -1}}\,,
\end{equation}
\begin{equation}\label{profile}\begin{aligned}
    \rho(r)=\Tilde{\rho}_0\left[\left(1-\frac{2 M}{r}\right)^{-\frac{1}{2}}-1\right]^{\frac{1}{\gamma -1}}+\frac{\Tilde{\rho}_0}{\gamma}\left[\left(1-\frac{2 M}{r}\right)^{-\frac{1}{2}}-1\right]^{\frac{\gamma }{\gamma -1}}\,.
\end{aligned}\end{equation}

\begin{figure*}
    \subfloat{
            \includegraphics[width=0.45\linewidth]{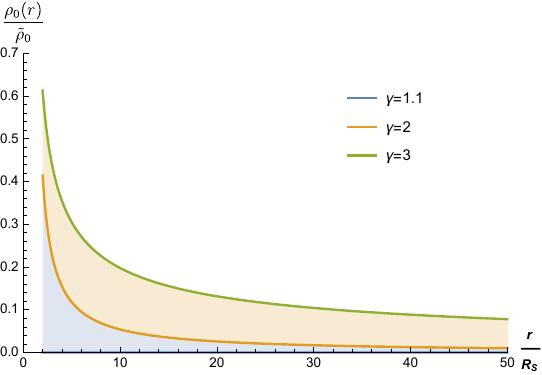}
       }
        \hfill
    \subfloat{
            \includegraphics[width=0.45\linewidth]{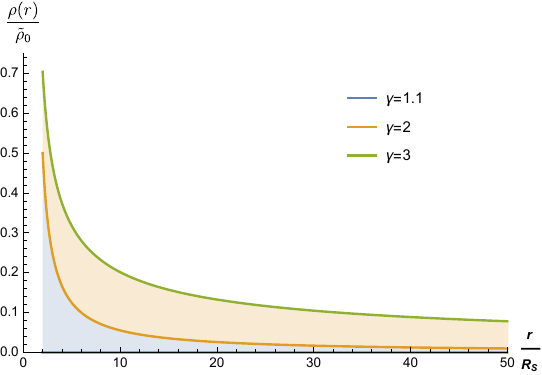}
       }

    \caption{The rest-mass density profile $\rho_0(r)$ and the total energy density profile $\rho(r)$ for different adiabatic indices $\gamma$. 
    When $r<4M$, the density profiles of DM vanish, where $R_S=2M$ is the Schwarzschild radius. Here we use the units such that $c=G=2M=1$.}
    \label{fig:profile}
\end{figure*}

\begin{figure*}
        \includegraphics[width=0.45\linewidth]{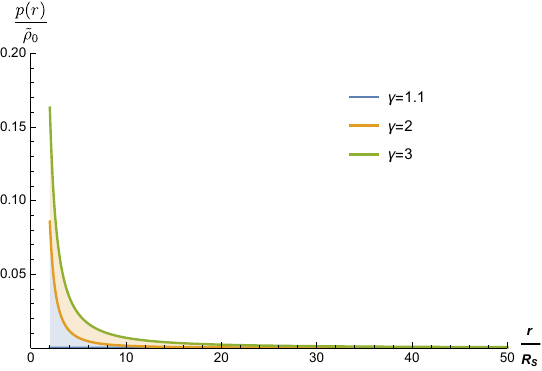}
         \caption{The pressure for different adiabatic indices $\gamma$, where $R_S=2M$ is the Schwarzschild radius. Here we use the units such that $c=G=2M=1$.}
    \label{fig:pressure}
\end{figure*}

Based on~\cite{PhysRevD.88.063522,PhysRevD.106.044027}, the density of DM spike peaks near $r\gtrsim 8M$ and vanishes below $r\sim 8M$.
The location of the spike decreases approximately to $r\sim 4M$ for cases with relativistic modification, attributed to DM particles annihilation or falling into the BH.
Given the above considerations, we choose a cutoff radius $r=4M$ in the density profiles governed by Eq.~\eqref{ad}.
The resultant profiles for different regions are listed in Table~\ref{Tab:region}.
Generally speaking, all three quantities $\rho_0(r)$, $\rho(r)$ and $p(r)$ become larger with increasing $\gamma$, as shown in Figs.~\ref{fig:profile} and~\ref{fig:pressure}.
As expected, the total energy density is larger than the rest-mass energy density for given $\gamma$, as can be inferred from Fig.~\ref{fig:profile}.
It is noted that although the choice of the location of the cutoff impacts spacetime and corresponding GWs, it does not cause significant deviation for the low-lying QNMs~\cite{PhysRevD.108.024070}.
Since our discussion focuses on the effects of various EOSs, we have chosen a unique cutoff value among different scenarios.

\subsection{Modified black hole metric}\label{sec:gauge}

Now we turn to discuss the corresponding modifications to the black hole metric by making use of the three obtained unknowns $\left[\rho(r), p(r),\rho_0(r)\right]$, which govern the behavior of DM spike associated with Bondi accretion. 
The impact on the spacetime metric is evaluated by employing two remaining TOV equations, namely, Eqs.~\eqref{TOV1} and~\eqref{TOV2} in terms of the metric functions $\left[f(r), g(r)\right]$. 

We first consider Eq.~\eqref{TOV1}. 
By substituting Eq.~\eqref{profile} into Eq.~\eqref{TOV1} and integrating $r$ over the range $r\in (4M,\infty)$, we have
\begin{equation}\label{gsol}
    \begin{aligned}
        g(r)&=1-\frac{2 M}{r}-\frac{8 \pi  G }{r}\int _{4 M}^rR^2 \rho (R)dR\\
        &=1-\frac{2 M}{r}-\frac{ 8 \pi  G  M^3 \Tilde{\rho}_0}{\gamma  r}[g_2(r)-g_1]-\frac{ 8 \pi  G  M^3 \Tilde{\rho}_0}{r}[g_4(r)-g_3]\,,
    \end{aligned}
\end{equation}
where one has chosen the appropriate constant of integration so that $g(4M)=1-\frac{2M}{r}$.
The two radial functions $g_2(r)$ and $g_4(r)$ come from the integration of the two terms of $\rho(r)$ in Eq.~\eqref{profile}.
They are defined as
\begin{equation}
    \begin{aligned}
        g_2(r)=\frac{1}{M^3}\int r^2\left[\left(1-\frac{2 M}{r}\right)^{-\frac{1}{2}}-1\right]^{\frac{\gamma }{\gamma -1}} dr\,,
    \end{aligned}
\end{equation}
and
\begin{equation}
    \begin{aligned}
        g_4(r)=\frac{1}{M^3}\int r^2\left[\left(1-\frac{2 M}{r}\right)^{-\frac{1}{2}}-1\right]^{\frac{1}{\gamma -1}} dr\,,
    \end{aligned}
\end{equation}
whose specific forms can be found in Appendix~\ref{app:gaugeg}.
Also, $g_1\equiv g_2(4M)$ and $g_3\equiv g_4(4M)$. 

To derive the form of $f(r)$, we consider Eq.~\eqref{TOV2}.
We note that the effective density $M^2\Tilde{\rho}_0\ll 1$, and therefore we neglect the second and higher order terms of $\Tilde{\rho}_0$ and obtain
\begin{equation}\label{Dlogf}
    \begin{aligned}
        \frac{d}{dr}\ln f &= \frac{f'(r)}{f(r)}=\frac{8 \pi  G r p(r)}{g(r)}+\frac{1}{r g(r)}-\frac{1}{r}\\
        &\sim\frac{2 M}{r (r-2 M)}+\frac{ 8 \pi  G \Tilde{\rho}_0 r^2 }{ r-2 M}\frac{\gamma -1}{\gamma}\left(\frac{1}{\sqrt{1-\frac{2 M}{r}}}-1\right)^{\frac{\gamma }{\gamma -1}}-\frac{8 \pi  G M^3 \Tilde{\rho}_0}{(r-2 M)^2}\left(\frac{g_1}{\gamma }+g_3\right)+\frac{8 \pi  G M^3 \Tilde{\rho}_0}{(r-2 M)^2} \left[\frac{g_2(r)}{\gamma }+g_4(r)\right]\,.
    \end{aligned}
\end{equation}
After integrating over $r\in (4M,\infty)$ and again neglecting the second and higher order terms of $\Tilde{\rho}_0$ as well as the terms of a higher order than $\left(1-\frac{1}{\sqrt{1-\frac{2 M}{r}}}\right)^{k_{\text{max}}}$, we find the following result
\begin{equation}\label{fsol}
    \begin{aligned}
        &f(r)=1-\frac{2M}{r}+8 \pi  G M^2 \Tilde{\rho}_0\left(1-\frac{2 M}{r}\right) \sum_{i=1}^6\left[f_i(r)-f_i(4M)\right]\,,
    \end{aligned}
\end{equation}
where $f_1(r)$ and $f_2(r)$ come from the integration of the first two terms in Eq.~\eqref{Dlogf}, $f_3(r)$ and $f_5(r)$ come from the integration of the non-hyper-geometric-function terms of $g_2(r)$ and $g_4(r)$, while $f_4(r)$ and $f_6(r)$ come from the integration of the hyper-geometric-function terms of $g_2(r)$ and $g_4(r)$. 
The specific forms of these terms are also relegated to Appendix~\ref{app:gaugef}.

The formalism in Appendixes~\ref{app:gaugeg} and \ref{app:gaugef} seem to indicate divergent terms in some special cases characterized by specific values of $\gamma$, such as $\gamma=1.5,2$.
Fortunately, as discussed in Appendix~\ref{app:gamma2}, such divergence always cancels out in pairs and the resulting expression remains well-defined analytically.
However, numerically, the presence of divergent terms does bring certain challenges.
Given the discussion presented in Appendix~\ref{app:gamma2}, it is proposed that the values of $\gamma=1.5$ and $\gamma=2$ be replaced with $\gamma=1.5+10^{-10}$ and $\gamma=2+10^{-10}$ respectively, throughout the numerical process in Sec.~\ref{sec:Method} and \ref{sec:result}.

It is noted that the above results are only associated with the region $r\in (4 M, \infty)$. 
For $r\in (2M,4 M)$ and $r\rightarrow \infty$, we have $\tilde\rho_0=0$, and the DM densities vanish with $\rho(r)=0$ as shown in Fig.~\ref{fig:profile}.
Subsequently, the spacetime falls back to the Schwarzschild case with $f(r)=g(r)=1-\frac{2M}{r}$. Tab.~\ref{Tab:region} enumerates different scenarios for both regions.

\begin{table}
    \caption{A summary of the relevant physical quantities and master equation for the QNMs in the two regions. 
    $\rho_0(r)$, $p(r)$, and $\rho(r)$ are the rest-mass density profile, pressure, and total energy density profile of DM. 
    $f(r)$ and $g(r)$ are the metric functions defined in Eq.\eqref{metric} and $r_*$ is the tortoise coordinate.}
    \begin{ruledtabular}
    \begin{tabular}{ccc}
    & $r\in(2M,4 M)$ & $r\in(4 M, \infty)$\\
    \hline 
    $\rho_0(r)$ & $0$ & Eq.~\eqref{rho0}\\
    $p(r)$ & $0$ & Eq.~\eqref{pressure}\\
    $\rho(r)$ & $0$ & Eq.~\eqref{profile}\\
    $f(r)$ & $1-\frac{2M}{r}$ & Eq.~\eqref{fsol}\\
    $g(r)$ & $1-\frac{2M}{r}$ & Eq.~\eqref{gsol}\\
    $r_*$ & Eq.~\eqref{tortoSch} & Eq.~\eqref{tortoAxial}\\
    master equation & Eq.~\eqref{QNMsRW} & Eq.~\eqref{QNMsAxial}
    \end{tabular}
    \end{ruledtabular}
    \label{Tab:region}
\end{table}

\section{Axial Perturbations of Schwarzschild-like black holes}\label{sec:QNMeqn}

This section examines the gravitational perturbations in the modified Schwarzschild-like background metric given by Eq.~\eqref{metric}. 
We shall focus on the axial perturbation in Sec.~\ref{sec:Axial}.
The master equations in different regions are derived and given in Tab.~\ref{Tab:region}. 
Notably, the presence of a discontinuity in the density profile results in different master equations for the QNMs, as shown in Figs.~\ref{fig:profile} and~\ref{fig:VAxial}.
The numerical method tailored for such a scenario will be discussed in Sec.~\ref{sec:Method}.

Typically, the perturbations can arise from the injection of gravitational waves or the infalling of a particle into the BHs~\cite{1988sfbh.book.....F, SchFallingParticle}, in addition to more extreme events such as the remnants of a binary BH merger. 
Such a perturbed metric can be described by:
\begin{equation}\label{LinearMetric}
    g_{\mu \nu}=\mathring{g}_{\mu \nu}+h_{\mu \nu}\,,
\end{equation}
where $\mathring{g}_{\mu \nu}$ is the metric of background spacetime given by Eq.~\eqref{metric} and Tab.~\ref{Tab:region}.
On the other hand, $h_{\mu \nu}$ represents the linear perturbation term. 
In deriving the master equation, the contributions owing to higher-order perturbations will be neglected.

The background spacetime $\mathcal{M}^4(t,r,\theta,\phi)$ possesses static spherical symmetry.
This implies that it can be expressed as the direct product of a 2-dimension Lorentzian manifold $M^2(t,r)$ and a 2-dimension unit sphere surface manifold $S^2(\theta,\phi)$.
Under the above symmetries, the metric perturbation $h_{\mu\nu}$ can be decomposed into various multipoles that will evolve independently in time according to the little group representation.
In particular, they can be classified into axial (odd) and polar (even) parity ones, described by~\cite{Thompson_2017, Nagar_2005}:
    \begin{align}
       & h_{\mu \nu}=\sum_{\ell=0}^{\infty}\sum_{m=-\ell}^{m=\ell}\left[\left(h_{\mu \nu}^{\ell m}\right)^{(\mathrm{axial})}+\left(h_{\mu \nu}^{\ell m}\right)^{(\mathrm{polar})}\right]\,,
    \end{align}
where, $\ell$ and $m$ are the integers from the separation of $\theta$ and $\phi$ respectively.
Regarding the Regge-Wheeler (RW) gauge, the axial perturbations can be parameterized as~\cite{PhysRevD.1.2870, Berti_2009}:
    \begin{align}
      & \left(h_{\mu \nu}^{\ell m}\right)^{(\mathrm{axial})}=\left(\sin \theta \frac{\partial Y_{\ell 0}(\theta)}{\partial \theta}\right)\mathrm{e}^{i\omega t}\epsilon\cdot\nonumber\\
      &\left(\begin{array}{cccc}
        0 & 0 & 0 & h_0(r) \\
        0 & 0 & 0 & h_1(r) \\
        0 & 0 & 0 & 0 \\
        h_0(r) & h_1(r) & 0 & 0
        \end{array}\right)\,.\label{axialCoe}
    \end{align}
Furthermore, that of the polar perturbations is:
    \begin{align}
        &\left(h_{\mu \nu}^{\ell m}\right)^{(\mathrm{polar})}=Y_{\ell 0}(\theta)\mathrm{e}^{i\omega t}\epsilon\cdot\nonumber\\
        &\left(\begin{array}{cccc}
        H_0(r) (1-\frac{2M}{r}) & H_1(r) & 0 & 0 \\
        H_1(r) & \frac{H_2(r)}{1-\frac{2M}{r}} & 0 & 0 \\
        0 & 0 & r^2 K(r) & 0 \\
        0 & 0 & 0 & r^2 K(r) \sin ^2 \theta
        \end{array}\right)\,,\label{polarCoe}
    \end{align}
where $\left| \epsilon \right| \ll 1$ is a real number to quantify the magnitude of perturbations proposed by~\cite{PhysRevD.104.124082}.
The eigenfrequency $\omega$ comes from separating the variable $t$, corresponding to the quasi-normal modes (QNMs) of the BHs.
The angular sector of the waveform is governed by the spherical harmonics $Y_{\ell m}$ with $m=0$ owing to the spherical symmetry.

It is significant to note that the axial gravitational perturbation is intrinsically decoupled from any scalar field.
Therefore, the perturbations of dark matter can be largely neglected, leading to mathematical simplification.
The resulting perturbation equation for the axial perturbations can then be derived, which will be given shortly in Sec.\ref{sec:Axial}.
Nevertheless, such a treatment is not valid for polar perturbations since they are likely to be coupled with scalar degrees of freedom, particularly DM.
As a result, deriving the perturbation equations for the polar case and computation of QNMs present a rather challenging problem~\cite{PhysRevLett.129.241103}.
In light of the above consideration, the primary objective of this paper is to concentrate solely on the axial perturbations.

\subsection{The master equation for axial gravitational perturbations}\label{sec:Axial}

As discussed above, per Refs.~\cite{liu2023gauge, PhysRevD.104.124082, PhysRevLett.129.241103}, the axial perturbations are decoupled from those in the DM.
Therefore, in what follows, we will concentrate on the axial perturbations of the metric.
To this end, we substitute Eq.~\eqref{axialCoe} into the Einstein equations $G_{\mu\nu}=8 \pi T_{\mu\nu}$ to derive the master equation.
For the spacetime region merged in the DM spike, the axial gravitational perturbations are governed by the following equation
\begin{equation}\label{QNMsAxial}
    \left[\frac{\partial^2}{\partial r_*^2}+\omega^2-V_{\text{axial}}(r)\right] \Psi(r)=0\,,
\end{equation}
where $r_*$ is the tortoise coordinate defined by
\begin{equation}\label{tortoAxial}
    d r_*=\frac{d r}{\sqrt{f(r) g(r)}}\,,
\end{equation}
and the effective potential reads
    \begin{align}
    V_{\text{axial}}(r)=\frac{r f^{\prime}(r) g^{\prime}(r)+g(r)\left[f^{\prime}(r)+2 r f^{\prime \prime}(r)\right]}{2 r}
    -\frac{g(r) f^{\prime}(r)^2}{2 f(r)}+\frac{f(r)\left[r g^{\prime}(r)+4 g(r)+2\left(\ell^2+\ell-2\right)\right]}{2 r^2}\,,
    \end{align}
where $g(r)$ and $f(r)$ are the metric functions given by Eqs.~\eqref{gsol} and~\eqref{fsol}.
When $r\leq 4 M$ or $\tilde\rho_0=0$, it falls back to the Schwarzschild case (see Tab.~\ref{Tab:region}) and the corresponding master equation is simplified to read
    \begin{equation}\label{QNMsRW}
    \left\{\frac{\partial^2}{\partial r_*^2}+\omega^2-\left(1-\frac{2M}{r}\right)\left[\frac{\ell(\ell+1)}{r^2}-\frac{6M}{r^3}\right]\right\} \Psi=0\,,
    \end{equation}
with $r_*$ given by:
    \begin{equation}\label{tortoSch}
    \mathrm{d} r_*=\left(1-\frac{2M}{r}\right)^{-1}\mathrm{d} r\,,
    \end{equation}
which is nothing but the well-known Regge–Wheeler–Zerilli equation~\cite{RW}.

\subsection{Matrix method for quasinormal modes in potential with discontinuity}\label{sec:Method}

\begin{figure*}
    \includegraphics[width=0.5\linewidth]{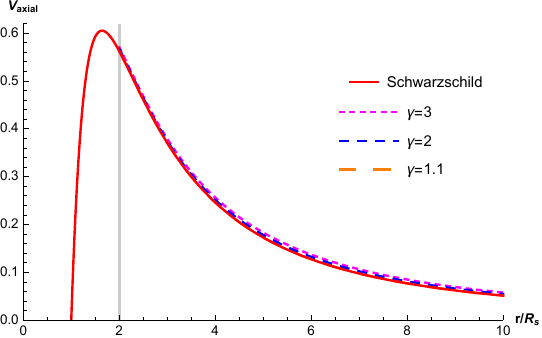}
    \caption{A comparison of the effective potentials for axial gravitational perturbations with and without DM. 
    In order to illustrate the difference, we assume $\tilde\rho_0=10^{-3}$. 
    When $r<4 M$, the DM vanishes, and the spacetime metric restores the form of the Schwarzschild BH, causing a discontinuity at $r=4 M$ where $R_S=2M$ is the Schwarzschild radius. 
    Here we adopt the units such that $c=G=2M=1$.}
    \label{fig:VAxial}
\end{figure*}

The master equations obtained above are also presented in Tab.~\ref{Tab:region}.
It is evident that a discontinuity occurs at $r=4 M$ when $\tilde\rho_0\neq 0$, as shown in Fig.~\ref{fig:VAxial}.
As known in the literature~\cite{agr-qnm-35, agr-qnm-36, agr-qnm-50, agr-qnm-lq-03, agr-qnm-echoes-20} that such a discontinuity entails non-trivial implications to the BH perturbation theory.
Meanwhile, several traditional methods for the BH QNMs, such as the WKB approximation~\cite{schutz1985black, PhysRevD.35.3621, PhysRevD.68.024018}, cannot be directly applied to the problem.
In this regard, we employ the modified matrix method to address the problem~\cite{Lin_2017,lin2017matrix,https://doi.org/10.48550/arxiv.2209.11612,Shen_2022,PhysRevD.107.124002}.

We proceed to discuss the boundary conditions for the master equation.
The bound of the relevant region consists of the horizon $r= 2M$ and spatial infinity $r\to \infty$, identical to the Schwartzchild case.
At these points, the asymptotical forms of the wave functions satisfy $\mathrm{e}^{-i\omega r_*}$ and $\mathrm{e}^{i\omega r_*}$, corresponding to the ingoing waves near the horizon and outgoing waves at spatial infinity, respectively~\cite{Berti_2009,https://doi.org/10.48550/arxiv.2212.00747}.
In the asymptotical regions, the tortoise coordinates $r_*$ defined by Eq.~\eqref{tortoSch} can be expressed explicitly as $r_*=r+2M \ln(r-2M)$.
Therefore, near the boundaries, the waveforms are governed by the following forms
    \begin{equation}\label{BC1}
        \Psi\rightarrow\begin{cases}
         \mathrm{e}^{-i\omega r}(r-2M)^{-2 i M \omega}  & r\rightarrow 2M  \\
            \mathrm{e}^{i \omega r} (r-2 M)^{2 i M \omega } & r\rightarrow\infty
            \\\end{cases}\, ,
    \end{equation}
which is asymptotically accurate up to an irrelevant constant.

We then introduce the transform of the wave function by the substitution
    \begin{equation}\label{changeL}
        \Psi(r)\equiv\mathrm{e}^{-i\omega r}(r-2M)^{-2 i M \omega} L(r)\,,
    \end{equation}
and
    \begin{equation}\label{coorL}
        y=\frac{r-2M}{2M}\,,
    \end{equation}
for the region $r\in (2M,4 M)$, and 
    \begin{equation}\label{changeR}
        \Psi(r)\equiv\mathrm{e}^{i \omega r} (r-2 M)^{2 i M \omega } R(r)\,,
    \end{equation}
and
    \begin{equation}\label{coorR}
        z=1-\frac{4   M}{r}\,,
    \end{equation}
for $r\in (4 M,\infty)$.
The transform introduced in Eqs.~\eqref{changeL} and~\eqref{changeR} effectively factors out the known asymptotical behavior of the wave functions at the boundaries.
In the meanwhile, Eq.~\eqref{coorL} and Eq.~\eqref{coorR} map $r\in (2M,4 M)$ and $r\in (4 M, \infty)$ to $[0,1]$ respectively.

The master equations given in Tab.~\ref{Tab:region} can be reformulated to read
    \begin{equation}
        A_2(y) L^{\prime\prime}(y) + A_1(y) L^{\prime}(y)+ A_0(y) L(y)=0\,,
    \end{equation}
    \begin{equation}
        B_2(z) R^{\prime\prime}(z) + B_1(z) R^{\prime}(z)+ B_0(z) R(z)=0\,,
    \end{equation}
where the coefficients $A_2$, $A_1$, $A_0$, $B_2$, $B_1$, and $B_0$ are functions of the variables $\omega$, $y$ (or $z$), and the DM parameters $\tilde\rho_0$ and $\gamma$.
Their specific forms are governed by the underlying master equation.

The boundary conditions given in Eq.~\eqref{BC1} can be rewritten as
\begin{equation}
    L(y=0)=const,\quad R(z=1)=const\,.
\end{equation}

For convenience, one further introduces
\begin{equation}
    \tilde{L}(y)\equiv y L(y)\,,
\end{equation}
\begin{equation}
    \tilde{R}(z)\equiv (1-z) R(z)\,,
\end{equation}
and the boundary conditions can be transformed into the form
\begin{equation}\label{BC2}
   \tilde{L}(z=0)=\tilde{R}(z=1)=0\,.
\end{equation}
The corresponding master equations now become
    \begin{equation}\label{eqnL}
        \tilde{A}_2(y) \tilde{L}^{\prime\prime}(y) + \tilde{A}_1(y) \tilde{L}^{\prime}(y)+ \tilde{A}_0(y) \tilde{L}(y)=0\,,
    \end{equation}
    \begin{equation}\label{eqnR}
        \tilde{B}_2(z) \tilde{R}^{\prime\prime}(z) + \tilde{B}_1(z) \tilde{R}^{\prime}(z)+ \tilde{B}_0(z) \tilde{R}(z)=0\,,
    \end{equation}
with
\begin{equation}
\begin{aligned}
        \tilde{A}_0(y)&=y^2 A_0\left(y\right)-y A_1\left(y\right)+2 A_2\left(y\right)\,,\\
        \tilde{A}_1(y)&=y \left[y A_1\left(y\right)-2 A_2\left(y\right)\right]\,,\\
        \tilde{A}_2(y)&=y^2 A_2\left(y\right)\,,\\
        \tilde{B}_0(z)&=\left(z-1\right){}^2 B_0\left(z\right)-\left(z-1\right) B_1\left(z\right)+2 B_2\left(z\right)\,,\\
        \tilde{B}_1(z)&=\left(z-1\right) \left[\left(z-1\right) B_1\left(z\right)-2 B_2\left(z\right)\right]\,,\\
        \tilde{B}_2(z)&=\left(z-1\right){}^2 B_2\left(z\right) \,.
\end{aligned}
\end{equation}

We proceed to address the discontinuity occurring at $r=r_c\equiv4 M$.
Such discontinuity in the metric must be in accordance with Israel's junction condition~\cite{israel1966nuovo}.
The wave functions, on the other hand, are connected through the requirement of vanishing Wronskian~\cite{PhysRevD.59.044034, https://doi.org/10.48550/arxiv.2209.11612, Shen_2022}:
\begin{equation}
    \Psi'(r=r_c^-)\Psi(r=r_c^+)-\Psi(r=r_c^-)\Psi'(r=r_c^+)=0\,,
\end{equation}
where $r=r_c^-$ and $r=r_c^+$ approach the discontinuity from the left and right sides, respectively.
The ratio coefficient $\kappa$ is subsequently defined as:
\begin{equation}
        \kappa=\frac{\Psi'(r=r_c^-)}{\Psi(r=r_c^-)}=\frac{\Psi'(r=r_c^+)}{\Psi(r=r_c^+)}\,.
\end{equation}
By substituting Eqs.~(\ref{changeL}-\ref{coorR}), the above expression can be reformulated as:
\begin{equation}\label{CCL}
\begin{aligned}
    y \tilde{L}'(y)+ \left[-2 \kappa  M y-2 i M (y+1) \omega -1\right]\tilde{L}(y)=0\,,
\end{aligned}
\end{equation}
\begin{equation}\label{CCR}
\begin{aligned}
      (z+1) (z-1)^2 \tilde{R}'(z)+ \left[-(z+1) (4 \kappa  M+z-1)+8 i M \omega \right]\tilde{R}(z)=0\,.
\end{aligned}
\end{equation}
Eqs.~\eqref{CCL} and~\eqref{CCR} furnish the connection conditions for the waveforms.

The matrix method algorithm for our specific case is outlined as follows:
\begin{enumerate}

    \item According to \cite{Lin_2017}, it is possible to discretize any coordinate $x\in [0,1]$ into a set of $N$ points denoted as $x_1,x_2,\cdots,x_{N}$.
    The use of Taylor expansion allows for the representation of a function, along with its first-order derivatives up to its $N$th-order derivatives, in the form of $N\times N$ matrices at each point.
    We relegate further details to Ref.~\cite{Lin_2017}, where a public version of the code is published in the arXiv website~\footnote{A public version of the code can be found via the link \url{https://arxiv.org/abs/1610.08135}.}.
    
    \item By substituting the matrices of the functions, first-order derivatives, and second-order derivatives obtained above, Eqs.~\eqref{eqnL} and~\eqref{eqnR} can be reformulated as two matrix equations, namely $\overline{\mathcal{M}}_L\mathcal{L}=\overline{\mathcal{M}}_R\mathcal{R}=0$.
    Here, $\overline{\mathcal{M}}_L$ and $\overline{\mathcal{M}}_R$ represent matrics of dimensions $N_L\times N_L$ and $N_R\times N_R$ respectively, which are solely associated to the variable $\omega$.
    In the meanwhile, $\mathcal{L}=\left(L(x_1), \cdots, L(x_{N_L})\right)^T$ and $\mathcal{R}=\left(R(x_1), \cdots, R(x_{N_R})\right)^T$ are the values of functions at each point.
    
    \item We apply Eq.~\eqref{BC2} to replace the first line of $\overline{\mathcal{M}}_L$ and $N$th line of $\overline{\mathcal{M}}_R$, respectively.
    Additionally, we employ Eqs.\eqref{CCL} and~\eqref{CCR} to replace the $N$th line of $\overline{\mathcal{M}}_L$ and first line of $\overline{\mathcal{M}}_R$, respectively.
    The equations $\mathcal{M}_L\mathcal{L}=\mathcal{M}_R\mathcal{R}=0$ are derived, where $\mathcal{M}_L$ and $\mathcal{M}_R$ represent the modified matrices with respect to $\omega$ and $\kappa$.
    
    \item By solving the equations $\det (\mathcal{M}_L)=\det (\mathcal{M}_R)=0$, the QNMs $\omega$ can be obtained together with the corresponding ratio $\kappa$.
\end{enumerate}

\section{Numerical results}\label{sec:result}

This section will examine the numerical results of fundamental QNMs and their properties influenced by the DM spike.
Moreover, the focus of our discussion is the impacts of different EOSs.
In order to speed up the calculation process, we choose $N_L=24$, $N_R=12$ and $k_{\text{max}}=20$ as in \cite{PhysRevD.108.024070}, which has been shown to provide reliable results up to six significant digits and eight for fundamental QNMs.

There are two parameters in our model, namely the adiabatic index $\gamma$ and the effective density parameter $\tilde{\rho}_0$.
The parameter $\gamma$ identifies different EOSs and falls in the relevant range $\gamma\in[1,3]$, according to the studies performed in~\cite{10.1093/mnras/stab161}.
In the present work, a few values of $\gamma$ will be explored, namely, $\gamma=\frac{5}{3}$, $\gamma=2$, and $\gamma=3$, whose impact on QNMs is also the focus of our exploration.

Regarding $\tilde{\rho}_0$, it is fixed by the requirement to match asymptotically the resulting spacetime with other DM profiles at a large scale.
Specifically, the DM density and pressure profiles describe the distribution very close to the BH, substantially impacting the spacetime around the BH and the corresponding GWs.
On the other hand, the distribution far away from the BH is often neglected~\cite{Nampalliwar_2021, PhysRevD.108.024070}.
In practice, however, in order to associate our model parameters $\Tilde{\rho}_0$ and $\gamma$ with realistic scenarios, we match the asymptotical spacetime to those of other models at a large scale $r\gg R_{sp}$.
For example, one can compare the present approach to the well-known NFW profile~\cite{Navarro_1997}:
\begin{equation}
    \rho(r)=\rho_{\mathrm{NFW}} \frac{r_{\mathrm{NFW}}}{r\left(1+\frac{r}{r_{\mathrm{NFW}}}\right)^2}\,.
\end{equation}
where $\rho_{\mathrm{NFW}}$ and $r_{\mathrm{NFW}}$ are parameters depending on different BHs or galaxies, which can be obtained from numerical fitting as in \cite{10.1093/mnras/stz1698}.

To proceed, in Sec.~\ref{sec:SgrA}, we then examine the QNMs using parameters for realistic gravitational systems, namely, the supermassive BHs at the center of Milky Way and M87.
In practice, the parameters close to the BHs might differ significantly from those for the bulk of the galaxy, or they vary substantially for different BHs and galaxies.
Therefore, a wide range of parameters will be considered in Sec.~\ref{sec:QNMPara} to assess the underlying impact.
Moreover, Sec.\ref{sec:detection} will address the detectability regarding the deviations of the quasinormal frequencies owing to the presence of DM.

\subsection{The QNMs of the supermassive BHs in Milky Way and M87}\label{sec:SgrA}

We first explore a few potential observational sources using realistic parameters and analyze the resulting BH QNMs.
Due to the limitation of the existing ground-based GW detectors~\cite{PhysRevLett.129.111102}, we primarily focus on the future space-based GW detectors.
As mentioned in Sec.~\ref{sec:profile}, the numerical fitting from observational data provides the parameterizations of some well-known profiles, such as the NFW profile.
For our present approach, the model parameters are extracted by matching the resultant DM profile to the existing ones at a large scale, $r\gg R_{sp}$.

In \cite{Nampalliwar_2021}, the DM spike structure for Sgr $\text{A}^*$ BH at the center of the Milky Way galaxy was explored.
The initial DM density profile is characterized as $\rho_0\sim r^{-\gamma_0}$, while that in the spike region is given by~\cite{PhysRevD.102.083006}:
\begin{equation}
    \rho_{\mathrm{DM}}^{\mathrm{sp}}(r)=\rho_{\mathrm{sp}}\left(\frac{R_{\mathrm{sp}}}{r}\right)^{\gamma_{\mathrm{sp}}}.
\end{equation}
This profile is then matched to observable data at large scales for $\gamma_{\mathrm{sp}}=\frac{7}{3}$ ($\gamma_0=1$), where we can obtain the values of parameters.
Therefore, we choose $\gamma=2$ to have a similar power law form for $\rho_0$, and then match the density profiles at a large scale, approximately $r\sim R_{\mathrm{sp}}$.

The mass of Sgr $\text{A}^*$ BH is $M=4.1\times 10^6 M_\odot$, while the best-fit values for the parameters are $R_{\mathrm{sp}}=0.235\text{kpc}$ and $\rho_{\mathrm{sp}}=8.00 \times 10^{-23}\text{g cm}^{-3}$.
Using the cutoff at $r\sim4M$, the parameters of Sgr $\text{A}^*$ BH are limited to $R_{\mathrm{sp}}=0.235\text{kpc}$ and $\rho_{\mathrm{sp}}<2.37\times 10^{-18} \text{g cm}^{-3}$~\cite{Nampalliwar_2021,Daghigh_2022}.
By matching the profiles at $r\sim R_{\mathrm{sp}}$ and choosing the unit $c=G=2 M=1$, we obtain the effective density parameter $\tilde\rho_0\approx 1.04\times 10^{-17}$ for the best-fit value and $\tilde\rho_0\approx 3.09\times10^{-13}$ as its upper limit. 
We then proceed to calculate the corresponding fundamental modes for the two values of $\tilde\rho_0$, which, up to the numerical precision, turn out to be identical, $0.7473433640-0.1779242954 i$.
In other words, the difference is not distinguishable.
One note that the fundamental mode of the Schwarzschild BH is $0.7473433526-0.1779242884 i$, and the deviation is still too small to be observationally relevant in the near future.

We therefore conclude that in order to detect the DM spike with $\gamma=2$, one needs to look for a more massive source.
A well-known example is the central BH in M87, which possesses a mass of $M=6.4\times10^{9}M_\odot$.
In this case, the best-fit parameters for M87* are $R_{\mathrm{sp}}=4.26\text{kpc}$ and $\rho_{\mathrm{sp}}=2.12 \times 10^{-23}\text{g cm}^{-3}$~\cite{Daghigh_2022,PhysRevD.96.063008}.
We note the same order of magnitude is obtained for $\rho_{\mathrm{sp}}$ in the best-fit case, and we thus adopt the same upper limit for M87*, $\rho_{\mathrm{sp}}<2.37\times 10^{-18} \text{g cm}^{-3}$.
Subsequently, one finds the effective density parameters to be, respectively, $\tilde\rho_0\approx7.81\times10^{-14}$ and $\tilde\rho_0\approx8.74\times 10^{-9}$.
The corresponding fundamental modes are found to be $0.7473433640-0.1779242954 i$ and $0.7473432167-0.1779244060 i$.
For the upper limit, the deviation in the quasinormal frequency is of the order $10^{-7}$, which gives $\sim10^{-12}$Hz and $\sim0.01s$.
Unfortunately, the feasibility of capturing such insignificant deviation is not optimistic as it largely resides outside the scope of future space-borne GW programs.

Nonetheless, it is noticed that the above analysis indicates that the mass of the BH indeed has a substantial impact on the DM distribution in terms of the effective density parameter $\tilde\rho_0$.
The latter modifies the quasinormal frequency, which is potentially relevant for more significant gravitational systems.
BHs with larger masses are prone to be detected.
Besides, as elaborated below, if one considers the scenarios by varying $\gamma$, the resulting impact on QNMs can be more favorable.

\subsection{The Fundamental modes with varying parameters}\label{sec:QNMPara}

In this subsection, we consider a broader range of parameters and investigate the modifications to the QNMs.
In particular, we explore the effects of two parameters: $\gamma$ and $\tilde\rho_0$.
The parameter $\gamma$ characterizes the properties of the DM, particularly the EOS.
As discussed above, we will focus on the range $\gamma\in[1,3]$.
By definition, the effective density parameter $\tilde\rho_0$ is associated with the initial distribution of DM and the characteristics of the underlying BH, particularly the mass.
Although we have already considered a few realistic cases by adopting specifically optimized parameters in the preceding subsection, exploring the parameter space is important for three main reasons~\cite{PhysRevD.104.124082, ZHANG2022101078}:
\begin{itemize}
    \item (1) Typically, The DM parameters primarily reflect the bulk distribution of DM in the galaxy.
    This is because they are obtained through a fit to the corresponding density profiles, which in turn is inferred from the data of rotation curves in various galaxies~\cite{10.1093/mnras/stz1698}. 
    However, in the vicinity of the BH, these parameters are largely free owing to the significant impact of the latter.
    \item (2) The baryonic component plays a substantial role in the DM parameters close to the BH.
    \item (3) The DM parameters $\gamma$ and $\tilde\rho_0$ and the mass of the central BH vary for different gravitational systems.
    A suitable BH merged in some particular DM spike might be easier for detection.
\end{itemize}

\begin{figure*}
    \subfloat{
            \includegraphics[width=0.45\linewidth]{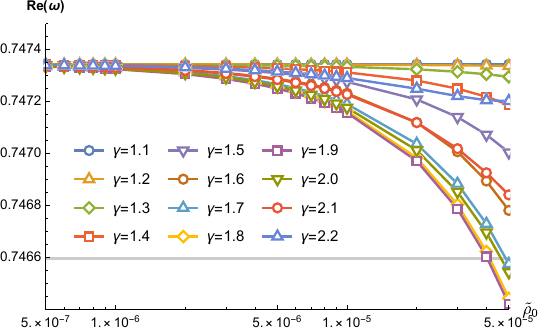}
       }
        \hfill
    \subfloat{
            \includegraphics[width=0.45\linewidth]{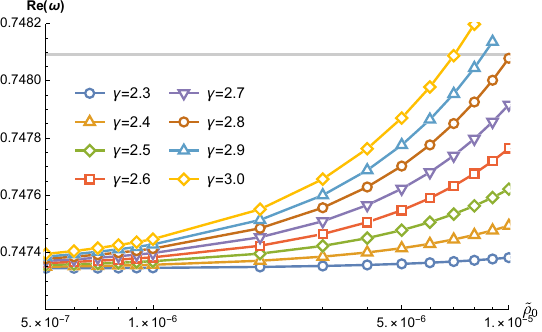}
       }
    \caption{The real parts of the QNMs as functions of $\Tilde{\rho}_0$ for different values of $\gamma$. 
    We adopt the units so that $c=G=2M=1$.}
    \label{fig:QNMrhoR}
\end{figure*}

\begin{figure*}
    \subfloat{
            \includegraphics[width=0.45\linewidth]{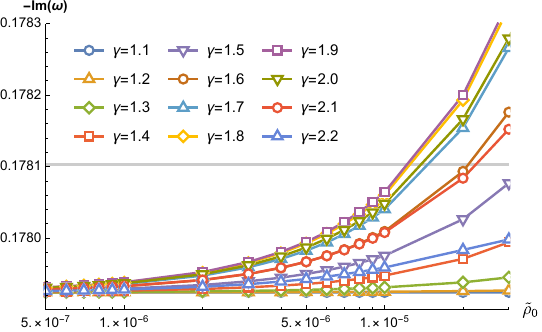}
       }
        \hfill
    \subfloat{
            \includegraphics[width=0.45\linewidth]{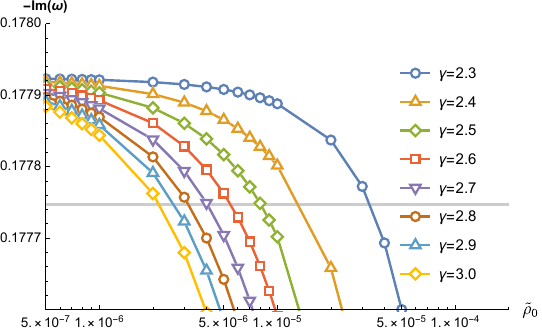}
       }

    \caption{The imaginary parts of the QNMs as functions of $\Tilde{\rho}_0$ for different values of $\gamma$. 
    We adopt the units so that $c=G=2M=1$.}
    \label{fig:QNMrhoI}
\end{figure*}

To this end, we will first evaluate the QNMs as functions of different $\gamma$ and $\tilde\rho_0$. 
As shown in Fig.~\ref{fig:QNMrhoR} and Fig.~\ref{fig:QNMrhoI}, for $\gamma \lesssim 2.2$, it is observed that both the real and the imaginary parts of QNMs decrease as $\tilde\rho_0$ increase, while they increase for $\gamma \gtrsim 2.2$.
In particular, as shown in Fig.~\ref{fig:QNMgamma}, there is a turning point near $\gamma\sim 2$.
For given $\tilde\rho_0$, as $\gamma$ increases, the real and imaginary parts of QNMs decrease and then increase when the turning point is passed.

Secondly, we focus on the impacts of a few values of $\gamma$ with specific physical interest.
$\gamma=\frac{5}{3}$ is roughly the watershed between the Newtonian and relativistic approaches, as elaborated in detail by Ref.~\cite{1952MNRAS.112..195B,10.1093/mnras/stab161}.
The EOSs with $\gamma\leq\frac{5}{3}$ correspond to soft ones, where the sound speeds are much less than the speed of light, and therefore, the Newtonian approach is expected to describe the DM accretion process well.
On the other hand, for stiff EOSs with $\gamma>\frac{5}{3}$, the Newtonian approach might cause the non-physical solutions, and the relativistic modification must be introduced.
However, as shown in Fig.~\ref{fig:QNMgamma}, it might be somewhat of a surprise not to observe any unusual behavior of QNMs around the value $\gamma=\frac{5}{3}$.
This can be understood since the relativistic modification has already been incorporated into our approach.

Additionally, the cases with $\gamma=2$ usually describe the interior of neutron stars, including the ultra-relativistic EOSs~\cite{10.1093/mnras/stab161}.
For the DM case, the case with $\gamma=2$ also describes the two-body interacting superfluid DM while that of $\gamma=3$ describes the three-body interacting superfluid DM~\cite{DeLuca_2023}.
Based on our result, $\gamma\sim 2$ produces the extreme values of QNMs for given $\tilde\rho_0$, while $\gamma\sim 3$ results in the most significant deviation on QNMs, as shown in Fig.~\ref{fig:QNMgamma}.

\subsection{Detectability on Space-based Detectors}\label{sec:detection}

We now turn to consider the detectability of the impacts on QMNs of DM spike, which is the so-called BH spectroscopy~\cite{PhysRevD.73.064030}.
The GW waveform during the ringdown process can be written as:
\begin{equation}\label{RDGW}
    h_{+}+i h_{\times}=\frac{M_z}{D_{\mathrm{L}}} \sum_{\ell m n} \mathcal{A}_{\ell m n} e^{i\left(f_{\ell m n} t+\phi_{\ell m n}\right)} e^{-t / \tau_{\ell m n}} S_{\ell m n}\,,
\end{equation}
where $M_z$, $D_L$, $\mathcal{A}_{\ell m n}$, $\phi_{\ell m n}$ represents the red-shifted BH mass, the luminosity distance to the source, the amplitude of the corresponding QNM, the phase coefficient respectively, and $S_{\ell m n}$ denotes the 2-spin-weighted spheroidal harmonics depending on the polar and azimuthal angles.
The actual ringdown waveform is the superposition of the axial and polar parity components.
The two parameters associated with QNMs are the GW frequency $f_{\ell m n}$ and the damping time $\tau_{\ell m n}$.
These parameters are defined as follows:
\begin{align}
&2 \pi f_{\ell m n}=\operatorname{Re}\left(\omega_{\ell m n}\right)\,,\\
&\tau_{\ell m n}=-\frac{1}{\operatorname{Im}\left(\omega_{\ell m n}\right)}\,,
\end{align}
where $\omega_{\ell m n}$ is the QNMs for given $(\ell,m,n)$. Here we consider only the fundamental mode with $(\ell,m,n)=(2,0,0)$ because it decays the slowest. 
Following~\cite{ZHANG2022101078}, the frequency and the damping time can be expanded as:
\begin{align}
&f_{\ell m n}=f_{\ell m n}^{\operatorname{Sch}}\left(1+\delta f_{\ell m n}\right)\,,\\
&\tau_{\ell m n}=\tau_{\ell m n}^{\mathrm{Sch}}\left(1+\delta \tau_{\ell m n}\right)\,,
\end{align}
where $f_{\ell m n}^{\operatorname{Sch}}$ and $\tau_{\ell m n}^{\mathrm{Sch}}$ are the QNM frequency and damping time for Schwarzschild case, while $\delta f_{\ell m n}$ and $\delta\tau_{\ell m n}$ are the corresponding relative deviations.

\begin{figure*}
    \subfloat{
            \includegraphics[width=0.45\linewidth]{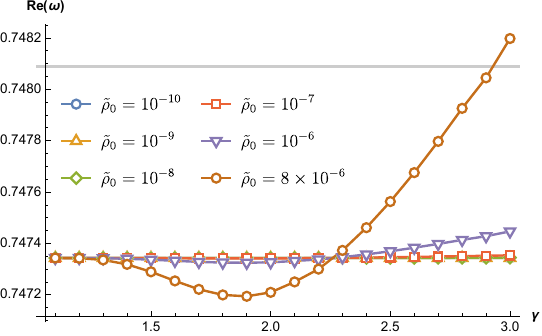}
       }
        \hfill
    \subfloat{
            \includegraphics[width=0.45\linewidth]{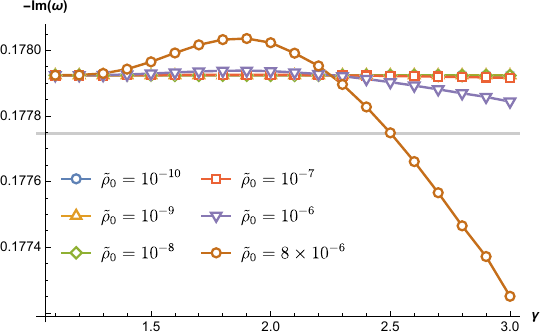}
       }

    \caption{The real and imaginary parts of axial QNMs as functions of $\gamma$ for different values of $\Tilde{\rho}_0$. The marked points are $\gamma=1.1,1.2,\cdots,3$. We adopt the units so that $c=G=2M=1$.}
    \label{fig:QNMgamma}
\end{figure*}

Given that the detection of ringdown signals by the ground-based GW detectors has not been successful~\cite{PhysRevLett.129.111102}, we turn to the future space-borne GW detectors, which possess a more promising sensitivity.
According to~\cite{PhysRevD.100.044036}, the relative deviation of fundamental modes $\delta\omega$ and $\delta\tau$ can be constrained within $0.0004\sim 0.002$ and $0.0005\sim 0.003$, with the ideal case of LISA-TianQin Joint detectors.
Thus, we assert that the relative deviation larger than $10^{-3}$ might be detected on space-based detectors.
When transformed into the deviations in quasinormal frequencies, we have the following conditions
\begin{equation}\label{detect1}
    \operatorname{Re}\left(\omega_{200}\right)>0.748091,\quad\text{or}\quad\operatorname{Re}\left(\omega_{200}\right)<0.746596\,,
\end{equation}
\begin{equation}\label{detect2}
         -\operatorname{Im}\left(\omega_{200}\right)>0.178102,\quad\text{or}\quad -\operatorname{Im}\left(\omega_{200}\right)<0.177747\,.
\end{equation}
The horizontal lines satisfying the conditions are indicated in Figs.~\ref{fig:QNMrhoR},~\ref{fig:QNMrhoI}, and~\ref{fig:QNMgamma}.
As shown in Figs.~\ref{fig:QNMrhoR} and~\ref{fig:QNMrhoI}, for sufficiently large values of $\tilde\rho_0$, the majority values of $\gamma$ give rise to detectable deviations in QNMs, except for $\gamma=1.1,1.2,1.3,1.4,2.2$.
Also, it is observed that various EOSs characterized by different values of $\gamma$ result in distinct behaviors of QNMs.
These results are rather inspiring and might be utilized in future GW detection and determination of the EOSs of DM, particularly in the context of EMRI systems.
The latter is largely attributed to their extensive observational period so that the relevant signals can be effectively accumulated over time.

\section{Further discussions and concluding remarks}

This paper explores the quasinormal frequencies of the GWs emitted by the perturbed Schwarzschild-like BHs merged in the DM spikes.
Particular attention is paid to the DM's EOS. 
By adopting Bondi's form, we consider the nonvanishing pressure of DM and explore its impact on the resulting QNMs. 
To this end, the density profile of the DM spike is derived regarding the adiabatic process in relativistic hydrodynamics.
The obtained profile is then furnished to the TOV equations in order to derive the modified BH metric. 
Subsequently, we obtain the master equation for the axial gravitational perturbations of the underlying metric, and the QNMs are evaluated using the modified matrix method.
We elaborate on the influence of the DM's EOS and discuss the detectability of the resulting modifications to the QNMs.

The key ingredients of the present study are as follows:
\begin{enumerate}
    \item For arbitrary adiabatic processes, Bondi's EOS is considered, which gives rise to nonvanishing pressure $p(r)$ in the DM.
    \item In the vicinity of a BH, the DM peaks near $r\gtrsim 4 M$ for cases with relativistic modification, while the distribution vanishes towards the inside due to annihilation or dropping into BH.
    Thus, we assume that the DM profile vanishes for $r\leq 4 M$ featuring a spike, and the highly nonlinear system of equations can be solved approximately in an iterative fashion.
    The density profiles $\rho_0(r)$ and $\rho(r)$ are derived and given in Eqs.~\eqref{rho0} and~\eqref{profile}. 
    These results are also shown in Figs.~\ref{fig:profile} and~\ref{fig:pressure}.
    The metric is given in Eqs.~\eqref{gsol} and~\eqref{fsol}.
    \item The spike in the DM profile divides the entire spatial domain into two different regions, as shown in Fig.~\ref{fig:profile}. 
    The resulting master equation, therefore, possesses a discontinuity as shown in Tab.~\ref{Tab:region} and Fig.~\ref{fig:VAxial}. 
    The latter is solved by adopting the modified matrix method, recently developed and tailored for such a scenario.
    \item Analysis for realistic scenarios such as the central BHs in Milky Way and M87 are considered in Sec.~\ref{sec:SgrA}. 
    The resulting deviations in the fundamental modes are too insignificant to be relevant for detection by the ongoing space-borne programs.
    \item For a given EOS or $\gamma$, when $\gamma \lesssim 2.2$, both the real and the imaginary parts of QNMs decrease as $\tilde\rho_0$ increases, while they both increase with $\tilde\rho_0$ for $\gamma \gtrsim 2.2$.
    The results are presented in Figs.~\ref{fig:QNMrhoR} and~\ref{fig:QNMrhoI}.
    \item For given $\tilde\rho_0$ there is a turning point near $\gamma\sim 2$. 
    On the one side, both real and imaginary parts of QNMs decrease as $\gamma$ increases, while on the other side, they both increase with increasing $\gamma$, as shown in Fig.~\ref{fig:QNMgamma}.
    \item The detectability of the deviations in QNMs is analyzed.
    Based on Eq.~\eqref{RDGW}, it is feasible for the ringdown waveforms to be detected by the future space-borne GW detectors in terms of relative deviations in the quasinormal frequency $\delta f_{lmn}$ and $\delta\tau_{lmn}$.
    By considering the detectable QNMs to satisfy the conditions given by Eqs.~\eqref{detect1} and~\eqref{detect2}, the threshold sensibility is indicated by the horizontal lines in Figs.~\ref{fig:QNMrhoR},~\ref{fig:QNMrhoI}, and~\ref{fig:QNMgamma}, in comparison with the obtained QNMs frequencies. 
    It is shown that different EOSs result in different detectability, and such features might be used to discriminate between DM models.
\end{enumerate}

To summarize, the presence of the DM spike owing to different EOSs essentially leaves detectable signatures on the resultant BH quasi-normal ringing in the GW waveform. 
Moreover, different DM EOSs give rise to different implications for the resulting GWs for a given DM density.
In particular, non-vanishing pressure plays a role and should be properly considered in pertinent studies.

DM accretion process often leads to a spike and, therefore, discontinuity in the matter distribution and effective potential, as shown in Table \ref{Tab:region}.
Such a discontinuity is understood to give birth to non-trivial effects.
Specifically, it was shown to have an intricate connection to the structural instability~\cite{PhysRevX.11.031003}.
In specific, even a minor ``ultraviolate'' perturbation, often expressed as a discontinuity in the effective potential~\cite{PhysRevD.103.024019}, significantly modifies high overtone QNMs leading to observational implications~\cite{PhysRevLett.128.111103}.
Furthermore, recent results~\cite{PhysRevD.106.084011} also indicate that it might destabilize the fundamental mode.
In this regard, further studies in the context of DM are also worthy topics.

Although the present paper only addresses spinless BHs, one may argue that the effect might be more substantial in rotating ones.
This is because the spin of a BH is likely to enhance the DM spike structure near BHs and the DM density near BHs~\cite{PhysRevD.96.083014}.
Therefore, more pronounced modifications to the QNMs and corresponding ringdown GWs are expected, which is favorable from an experimental perspective.

Last but not least, another intriguing topic involves the polar QNMs. 
As mentioned in Sec.~\ref{sec:QNMeqn}, such perturbations are likely to be coupled to the matter fields.
Moreover, this scenario was found to have a significant effect on EMRI GWs~\cite{PhysRevLett.129.241103}.
As a result, a more significant impact on ringdown GWs is expected, potentially facilitating detection.
We plan to address these topics in future exploration.


\acknowledgements


This work was supported in part by the National Key Research and Development Program of China Grant No. 2021YFC2203001 and in part by the NSFC (No.~11920101003, No.~12021003 and No.~12005016). Z. Cao was supported by ``the Interdiscipline Research Funds of Beijing Normal University" and CAS Project for Young Scientists in Basic Research YSBR-006. B. Sun is supported by the National Natural Science Foundation of China under Grants No.~12375046 and Beijing University of Agriculture Young Teachers Scientific Research Innovation Enhancement Program under Grants No.~QJKC-2023032.


\appendix

\section{Forms of the metric functions $g(r)$}\label{app:gaugeg}

Here we give the specific form of the metric functions used in Eq.~\eqref{gsol}. 
The function $g_2(r)$ comes from the integration of the second terms of $\rho(r)$ in Eq.~\eqref{profile}, which reads
\begin{equation}\begin{aligned}\label{g2}
    g_2(r)&=\frac{1}{24} \left(\frac{1}{\sqrt{1-\frac{2 M}{r}}}-1\right)^{\frac{1}{\gamma -1}}\left[\frac{4 \left(\frac{1}{\sqrt{1-\frac{2 M}{r}}}-1\right)^2}{\left(\frac{1}{\sqrt{1-\frac{2 M}{r}}}+1\right)^3}  \right.-\frac{6 \left(\sqrt{1-\frac{2 M}{r}}-1\right)^2}{\left(\sqrt{1-\frac{2 M}{r}}+1\right)^2}-60 (\gamma -1)\\
    &\left.+\frac{72 (\gamma -1)}{(\gamma -2) \left(\frac{1}{\sqrt{1-\frac{2 M}{r}}}-1\right)}+\frac{24 (\gamma -1)}{(2 \gamma -3) \left(\frac{1}{\sqrt{1-\frac{2 M}{r}}}-1\right)^2}-\frac{(11 \gamma -10) \left(\sqrt{1-\frac{2 M}{r}}-1\right)^2}{(\gamma -1) \left(\sqrt{1-\frac{2 M}{r}}+1\right)^2}\right]\\
    &+\frac{(5 \gamma -4) (12 \gamma -11)  }{96 (\gamma -1) (2 \gamma -1)}\left(\frac{1}{\sqrt{1-\frac{2 M}{r}}}-1\right)^{\frac{2 \gamma -1}{\gamma -1}}\, _2F_1\left[2,\frac{2 \gamma -1}{\gamma -1};\frac{3 \gamma -2}{\gamma -1};\frac{1}{2} \left(1-\frac{1}{\sqrt{1-\frac{2 M}{r}}}\right)\right]\,,
\end{aligned}\end{equation}
Moreover, $g_1\equiv g_2(4M)$ is given by
\begin{equation}\label{g1}\begin{aligned}
    g_1=g_2(4M)&=\frac{\left(\sqrt{2}-1\right)^{\frac{1}{\gamma -1}}}{96 (\gamma -1)}\left\{4 \left[\frac{72 \left(\sqrt{2}+1\right)}{\gamma -2}-320 \sqrt{2}+\frac{6 \left(2 \sqrt{2}+3\right)}{2 \gamma -3}+\gamma  \left(-60 \gamma +416 \sqrt{2}-225\right)+358\right]\right.\\
    &\left.-\frac{\left(2 \sqrt{2}-3\right) (5 \gamma -4) (12 \gamma -11) }{2 \gamma -1} \, _2F_1\left(2,\frac{2 \gamma -1}{\gamma -1};\frac{3 \gamma -2}{\gamma -1};\frac{1}{2}-\frac{1}{\sqrt{2}}\right)\right\}\,.
\end{aligned}\end{equation}
Similarly, $g_4(r)$ comes from the integration of the first terms of $\rho(r)$ in Eq.~\eqref{profile}, which reads
\begin{equation}\label{g4}
    \begin{aligned}
        g_4(r)&=\frac{\left(\frac{1}{\sqrt{1-\frac{2 M}{r}}}-1\right)^{\frac{\gamma }{\gamma -1}}}{6 \left(\frac{1}{\sqrt{1-\frac{2 M}{r}}}+1\right)^3} -\frac{2 \left(\frac{1}{\sqrt{1-\frac{2 M}{r}}}-1\right)^{\frac{\gamma }{\gamma -1}}}{3 \left(\frac{1}{\sqrt{1-\frac{2 M}{r}}}+1\right)^2}+\frac{3 (1-\gamma ) \left(\frac{1}{\sqrt{1-\frac{2 M}{r}}}-1\right)^{\frac{1}{\gamma -1}-2}}{3-2 \gamma }\\
        &+\frac{(1-\gamma ) \left(\frac{1}{\sqrt{1-\frac{2 M}{r}}}-1\right)^{\frac{1}{\gamma -1}-3}}{4-3 \gamma }+\frac{5 (1-\gamma ) \left(\frac{1}{\sqrt{1-\frac{2 M}{r}}}-1\right)^{\frac{1}{\gamma -1}-1}}{2 (2-\gamma )}-\frac{\left(\frac{1}{\sqrt{1-\frac{2 M}{r}}}-1\right)^{\frac{\gamma }{\gamma -1}}}{24 (\gamma -1) \left(\frac{1}{\sqrt{1-\frac{2 M}{r}}}+1\right)^2}\\
        &+\frac{\gamma  (44 \gamma -73)+30 }{96 (\gamma -1) \gamma }\left(\frac{1}{\sqrt{1-\frac{2 M}{r}}}-1\right)^{\frac{\gamma }{\gamma -1}}\, _2F_1\left[2,\frac{\gamma }{\gamma -1};2+\frac{1}{\gamma -1};\frac{1}{2} \left(1-\frac{1}{\sqrt{1-\frac{2 M}{r}}}\right)\right]\,,
    \end{aligned}    
\end{equation}
Moreover, $g_3\equiv g_4(4M)$ is given by
\begin{equation}\label{g3}
    \begin{aligned}
        g_3=&g_4(4M)=\frac{1}{96 (\gamma -1)}\left\{\frac{48 (\gamma -1)^2 \left(\gamma  \left(86 \sqrt{2} \gamma +112 \gamma -275 \sqrt{2}-363\right)+72 \left(3 \sqrt{2}+4\right)\right) \left(\sqrt{2}-1\right)^{\frac{1}{\gamma -1}}}{(\gamma -2) (2 \gamma -3) (3 \gamma -4)}\right.\\
        &+4 \left[\left(52 \sqrt{2}-76\right) \gamma -50 \sqrt{2}+73\right] \left(\sqrt{2}-1\right)^{\frac{\gamma }{\gamma -1}}\\
        &\left.+\frac{[\gamma  (44 \gamma -73)+30] \left(\sqrt{2}-1\right)^{\frac{\gamma }{\gamma -1}} }{\gamma }\, _2F_1\left(2,\frac{\gamma }{\gamma -1};2+\frac{1}{\gamma -1};\frac{1}{2}-\frac{1}{\sqrt{2}}\right)\right\}\,.
    \end{aligned}
\end{equation}
where ${ }_2 F_1(a, b ; c ; z)$ is the hyper geometric function defined as:
\begin{equation}\label{H-Gfun}
    { }_2 F_1(a, b ; c ; z)=\sum_{k=0}^{\infty} \frac{(a)_k(b)_k}{(c)_k} \frac{z^k}{k !}, \quad(a)_k=a(a+1) \cdots(a+k-1)\,.
\end{equation}

\section{Forms of the metric functions $f(r)$}\label{app:gaugef}

Here we give the specific form of the metric functions used in Eq.~\eqref{fsol}. 
The functions $f_1(r)$ and $f_2(r)$ come from the integrations of the first two terms in Eq.~\eqref{Dlogf}, namely,
\begin{equation}
    f_1(r)=\frac{1}{M^2}\frac{\gamma -1}{\gamma }\int \frac{ r^2}{ (r-2 M)}\left(\frac{1}{\sqrt{1-\frac{2 M}{r}}}-1\right)^{\frac{\gamma }{\gamma -1}} \, dr\,,
\end{equation}
\begin{equation}
    f_2(r)=-M\left(\frac{g_1}{\gamma }+g_3\right) \int \frac{1}{(r-2 M)^2} \, dr\,,
\end{equation}
The functions $f_3(r)$ and $f_5(r)$ come from the integrations of the non-hyper-geometric terms $g_2(r)$ and $g_4(r)$ in Eqs.~\eqref{g2} and~\eqref{g4}, which read
\begin{equation}\begin{aligned}
    f_3(r)&=\frac{M }{\gamma }\int \frac{1}{(r-2 M)^2}\\
    &\left\{g_2(r)-\frac{(5 \gamma -4) (12 \gamma -11)  }{96 (\gamma -1) (2 \gamma -1)}\left(\frac{1}{\sqrt{1-\frac{2 M}{r}}}-1\right)^{\frac{2 \gamma -1}{\gamma -1}}\, _2F_1\left[2,\frac{2 \gamma -1}{\gamma -1};\frac{3 \gamma -2}{\gamma -1};\frac{1}{2} \left(1-\frac{1}{\sqrt{1-\frac{2 M}{r}}}\right)\right]\right\} \, dr\,,
\end{aligned}\end{equation}
\begin{equation}
    \begin{aligned}
      f_5(r)&=M \int \frac{1}{(r-2 M)^2}\\
      &\left\{g_4(r)-\frac{\gamma  (44 \gamma -73)+30 }{96 (\gamma -1) \gamma }\left(\frac{1}{\sqrt{1-\frac{2 M}{r}}}-1\right)^{\frac{\gamma }{\gamma -1}}\, _2F_1\left[2,\frac{\gamma }{\gamma -1};2+\frac{1}{\gamma -1};\frac{1}{2} \left(1-\frac{1}{\sqrt{1-\frac{2 M}{r}}}\right)\right]\right\}\, dr\,,
    \end{aligned}
\end{equation}
On the other hand, $f_4(r)$ and $f_6(r)$ come from the integrations of the hyper-geometric terms of $g_2(r)$ and $g_4(r)$ in Eqs.~\eqref{g2} and~\eqref{g4}, which read
\begin{equation}\begin{aligned}
    f_4(r)&=\frac{M }{\gamma }\frac{(5 \gamma -4) (12 \gamma -11)  }{96 (\gamma -1) (2 \gamma -1)}\\
    &\int \frac{1}{(r-2 M)^2}\left\{\left(\frac{1}{\sqrt{1-\frac{2 M}{r}}}-1\right)^{\frac{2 \gamma -1}{\gamma -1}}\, _2F_1\left[2,\frac{2 \gamma -1}{\gamma -1};\frac{3 \gamma -2}{\gamma -1};\frac{1}{2} \left(1-\frac{1}{\sqrt{1-\frac{2 M}{r}}}\right)\right]\right\} \, dr\,,
\end{aligned}\end{equation}
\begin{equation}
    \begin{aligned}
      f_6(r)&=M \frac{\gamma  (44 \gamma -73)+30 }{96 (\gamma -1) \gamma } \\
      &\int \frac{1}{(r-2 M)^2}\left\{\left(\frac{1}{\sqrt{1-\frac{2 M}{r}}}-1\right)^{\frac{\gamma }{\gamma -1}}\, _2F_1\left[2,\frac{\gamma }{\gamma -1};2+\frac{1}{\gamma -1};\frac{1}{2} \left(1-\frac{1}{\sqrt{1-\frac{2 M}{r}}}\right)\right]\right\}\, dr\,.
    \end{aligned}
\end{equation}

Specifically, the function $f_1(r)$ and $f_2(r)$ are
\begin{equation}
    \begin{aligned}
        f_1(r)&=-\frac{1}{16 \gamma ^2}\left\{4 (\gamma -1) \gamma  \left(\frac{1}{\sqrt{1-\frac{2 M}{r}}}-1\right)^{\frac{1}{\gamma -1}} \left[14 (\gamma -1)-\frac{4 (\gamma -1)}{(\gamma -2) \left(\frac{1}{\sqrt{1-\frac{2 M}{r}}}-1\right)}+\frac{\left(\sqrt{1-\frac{2 M}{r}}-1\right)^2}{\left(\sqrt{1-\frac{2 M}{r}}+1\right)^2}\right] \right.\\
        &+64 (\gamma -1)^2 \left(\frac{1}{\sqrt{1-\frac{2 M}{r}}}-1\right)^{\frac{\gamma }{\gamma -1}} \, _2F_1\left(1,\frac{\gamma }{\gamma -1};2+\frac{1}{\gamma -1};1-\frac{1}{\sqrt{1-\frac{2 M}{r}}}\right)\\
        &+32 (\gamma -1) \gamma  \Gamma \left(2+\frac{1}{\gamma -1}\right) \left(\frac{1}{\sqrt{1-\frac{2 M}{r}}}-1\right)^{\frac{1}{\gamma -1}+2} \, _2\tilde{F}_1\left[1,2+\frac{1}{\gamma -1};3+\frac{1}{\gamma -1};\frac{1}{2} \left(1-\frac{1}{\sqrt{1-\frac{2 M}{r}}}\right)\right]\\
        &+64 (\gamma -1) \gamma  \Gamma \left(2+\frac{1}{\gamma -1}\right) \left(\frac{1}{\sqrt{1-\frac{2 M}{r}}}-1\right)^{\frac{1}{\gamma -1}+2} \, _2\tilde{F}_1\left(1,2+\frac{1}{\gamma -1};3+\frac{1}{\gamma -1};1-\frac{1}{\sqrt{1-\frac{2 M}{r}}}\right)\\
        &\left.-\gamma  (14 \gamma -13) \Gamma \left(2+\frac{1}{\gamma -1}\right) \left(\frac{1}{\sqrt{1-\frac{2 M}{r}}}-1\right)^{\frac{1}{\gamma -1}+2} \, _2\tilde{F}_1\left[2,2+\frac{1}{\gamma -1};3+\frac{1}{\gamma -1};\frac{1}{2} \left(1-\frac{1}{\sqrt{1-\frac{2 M}{r}}}\right)\right]\right\}\, ,
    \end{aligned}
\end{equation}
and
\begin{equation}
    \begin{aligned}
        f_2(r)&=\frac{M }{r-2 M}\left(\frac{g_1}{\gamma }+g_3\right)\, ,
    \end{aligned}
\end{equation}
where $\Gamma$ represents the Gamma function and $\, _2\tilde{F}_1\left(a,b;c;d\right)\equiv\, _2F_1\left(a,b,c,d\right)/ \Gamma(c)$ is the regularized hypergeometric function.

By performing the integration, $f_3(r)$ and $f_5(r)$ are given by:
\begin{equation}
    \begin{aligned}
        f_3(r)&=-\frac{3 (\gamma -1)^2}{(\gamma -2) \gamma ^2}\left(\frac{\gamma }{\frac{1}{\sqrt{1-\frac{2 M}{r}}}-1}+1\right) \left(\frac{1}{\sqrt{1-\frac{2 M}{r}}}-1\right)^{\frac{\gamma }{\gamma -1}}\\
        &+\frac{(1-\gamma ) (\gamma -1) }{\gamma  (2 \gamma -3)}\left[\frac{1}{(2-\gamma ) \left(\frac{1}{\sqrt{1-\frac{2 M}{r}}}-1\right)}+1\right]\left(\frac{1}{\sqrt{1-\frac{2 M}{r}}}-1\right)^{\frac{1}{\gamma -1}}\\
        &-\frac{5 (\gamma -1) }{2 \gamma }\left[\frac{(1-\gamma ) \left(\frac{1}{\sqrt{1-\frac{2 M}{r}}}-1\right)^{\frac{\gamma }{\gamma -1}}}{\gamma }-\frac{\left(\frac{1}{\sqrt{1-\frac{2 M}{r}}}-1\right)^{\frac{1}{\gamma -1}+2}}{\frac{1}{\gamma -1}+2}\right]\\
        &-\frac{1}{8 \gamma }\left(\frac{1}{\sqrt{1-\frac{2 M}{r}}}-1\right)^{\frac{1}{\gamma -1}+3} \left\{\frac{3-4 \gamma  }{3 \gamma -2}\, _2F_1\left[1,3+\frac{1}{\gamma -1};4+\frac{1}{\gamma -1};\frac{1}{2} \left(1-\frac{1}{\sqrt{1-\frac{2 M}{r}}}\right)\right]+\frac{2}{\frac{1}{\sqrt{1-\frac{2 M}{r}}}+1}\right\}\\
        &+\frac{1}{96 \gamma }\left(\frac{1}{\sqrt{1-\frac{2 M}{r}}}-1\right)^{\frac{1}{\gamma -1}+3} \left\{\frac{4-5 \gamma }{3 \gamma -2} \, _2F_1\left[2,3+\frac{1}{\gamma -1};4+\frac{1}{\gamma -1};\frac{1}{2} \left(1-\frac{1}{\sqrt{1-\frac{2 M}{r}}}\right)\right]+\frac{4}{\left(\frac{1}{\sqrt{1-\frac{2 M}{r}}}+1\right)^2}\right\}\, ,
    \end{aligned}
\end{equation}
and
\begin{equation}
    \begin{aligned}
        f_5(r)&=\frac{3 (\gamma -1)^2 }{3-2 \gamma }\left(\frac{1}{\sqrt{1-\frac{2 M}{r}}}-1\right)^{\frac{1}{\gamma -1}} \left[\frac{1}{(2-\gamma ) \left(\frac{1}{\sqrt{1-\frac{2 M}{r}}}-1\right)}+1\right]\\
        &+\frac{5 (1-\gamma ) (\gamma -1) }{2 (\gamma -2) \gamma }\left(\frac{1}{\sqrt{1-\frac{2 M}{r}}}-1\right)^{\frac{\gamma }{\gamma -1}} \left(\frac{\gamma }{\frac{1}{\sqrt{1-\frac{2 M}{r}}}-1}+1\right)\\
        &+\frac{(1-\gamma ) }{4-3 \gamma }\left(\frac{1}{\sqrt{1-\frac{2 M}{r}}}-1\right)^{\frac{1}{\gamma -1}-2} \left[\frac{\gamma -1}{2 \gamma -3}+\frac{1-\gamma  }{2-\gamma }\left(\frac{1}{\sqrt{1-\frac{2 M}{r}}}-1\right)\right]\\
        &-\frac{1}{6} \left(\frac{1}{\sqrt{1-\frac{2 M}{r}}}-1\right)^{\frac{1}{\gamma -1}+2}\left\{\frac{2-3 \gamma  }{2 \gamma -1}\, _2F_1\left[1,2+\frac{1}{\gamma -1};3+\frac{1}{\gamma -1};\frac{1}{2} \left(1-\frac{1}{\sqrt{1-\frac{2 M}{r}}}\right)\right]+\frac{2}{\frac{1}{\sqrt{1-\frac{2 M}{r}}}+1}\right\}\\
        &-\frac{1}{96 (\gamma -1)}\left(\frac{1}{\sqrt{1-\frac{2 M}{r}}}-1\right)^{\frac{1}{\gamma -1}+2}\left\{\frac{2-3 \gamma }{2 \gamma -1}\, _2F_1\left[1,2+\frac{1}{\gamma -1};3+\frac{1}{\gamma -1};\frac{1}{2} \left(1-\frac{1}{\sqrt{1-\frac{2 M}{r}}}\right)\right]+\frac{2}{\frac{1}{\sqrt{1-\frac{2 M}{r}}}+1}\right\}\\
        &+\frac{1}{96}\left(\frac{1}{\sqrt{1-\frac{2 M}{r}}}-1\right)^{\frac{1}{\gamma -1}+2}\left\{\frac{3-4 \gamma }{2 \gamma -1}\, _2F_1\left[2,2+\frac{1}{\gamma -1};3+\frac{1}{\gamma -1};\frac{1}{2} \left(1-\frac{1}{\sqrt{1-\frac{2 M}{r}}}\right)\right]+\frac{4}{\left(\frac{1}{\sqrt{1-\frac{2 M}{r}}}+1\right)^2}\right\}\, .
    \end{aligned}
\end{equation}

The functions $f_4(r)$ and $f_6(r)$ require the integration of the hyper-geometric terms of $g_2(r)$ and $g_4(r)$ in Eq.~\eqref{g2} and Eq.~\eqref{g4}.
Using Eq.~\eqref{H-Gfun}, the hypergeometric function can be expanded into series, and one may carry out the integration order by order.
Note that the integration interval is $r\in (4M,\infty)$, and $\left(1-\frac{1}{\sqrt{1-\frac{2 M}{r}}}\right)$ is a small quantity compared with $1$.
Therefore, any term of the order higher than $\left(1-\frac{1}{\sqrt{1-\frac{2 M}{r}}}\right)^{k_{\text{max}}}$ can be neglected.
Following this strategy, the specific forms of $f_4(r)$ and $f_6(r)$ are found to be
\begin{equation}
    \begin{aligned}
        f_4(r)&=-\frac{(5 \gamma -4) (12 \gamma -11)}{\gamma  (96 (\gamma -1) (2 \gamma -1))}\sum _{k=0}^{k_{\text{max}}} \frac{(-2)^{-k}}{k!} \frac{(2)_k \left(\frac{2 \gamma -1}{\gamma -1}\right)_k}{\left(\frac{3 \gamma -2}{\gamma -1}\right)_k} \left(\sqrt{\frac{1}{1-\frac{2 M}{r}}}-1\right)^{\frac{1}{\gamma -1}+k+3} \frac{1}{\frac{1}{\gamma -1}+k+3}+\frac{\sqrt{\frac{1}{1-\frac{2 M}{r}}}-1}{\frac{1}{\gamma -1}+k+4}\, ,
    \end{aligned}
\end{equation}
and
\begin{equation}
    \begin{aligned}
        f_6(r)=-\frac{\gamma  (44 \gamma -73)+30}{96 (\gamma -1) \gamma }\sum _{k=0}^{k_{\text{max}}} \frac{(-2)^{-k}}{k!} \frac{(2)_k \left(\frac{\gamma }{\gamma -1}\right)_k}{\left(2+\frac{1}{\gamma -1}\right)_k} \left(\sqrt{\frac{1}{1-\frac{2 M}{r}}}-1\right)^{\frac{1}{\gamma -1}+k+2} \frac{1}{\frac{1}{\gamma -1}+k+2}+\frac{\sqrt{\frac{1}{1-\frac{2 M}{r}}}-1}{\frac{1}{\gamma -1}+k+3}\, ,
    \end{aligned}
\end{equation}
where $(a)_k$ represents the Pochhammer symbol defined by:
\begin{equation}
    (a)_k=a(a+1) \cdots(a+k-1)\, .
\end{equation}

\section{The continuity of the metric functions near special values of $\gamma$}\label{app:gamma2}

It has been observed that certain exceptional scenarios, which entail specific values of $\gamma$, seem impossible within our formalism of the modified metric as given in Eqs.~\eqref{gsol}, \eqref{fsol} and Appendixes \ref{app:gaugeg}, \ref{app:gaugef}. For instance, when selecting $\gamma=2$, certain terms in Eqs.~\eqref{g2} and \eqref{g1} become divergent, which reads:
\begin{equation}\begin{aligned}\label{g22}
    {}_2g_2(r)&=\frac{1}{24} \left(\frac{1}{\sqrt{1-\frac{2 M}{r}}}-1\right)^{\frac{1}{\gamma -1}}\left[\frac{72 (\gamma -1)}{(\gamma -2) \left(\frac{1}{\sqrt{1-\frac{2 M}{r}}}-1\right)}\right]=\frac{3 (\gamma -1) }{\gamma -2}\left(\frac{1}{\sqrt{1-\frac{2 M}{r}}}-1\right)^{\frac{1}{\gamma -1}-1}\,,
\end{aligned}\end{equation}
and
\begin{equation}\label{g12}\begin{aligned}
    {}_2g_1={}_2g_2(4M)&=\frac{3 \left(\sqrt{2}-1\right)^{\frac{1}{\gamma -1}-1} (\gamma -1)}{\gamma -2}\,,
\end{aligned}\end{equation}
where we use the left subscript to denote the value $\gamma=2$. However, given that these divergent terms consistently appear in the form of differences as given in Eqs.~\eqref{gsol} and \eqref{fsol}, we apply the Taylor expansion of such terms and find:
\begin{equation}
    \begin{aligned}
        {}_2g_2(r)-{}_2g_1=\frac{3 \left(\sqrt{2}-1\right)^{\frac{3-2 \gamma }{\gamma -1}} }{4 \sqrt{2} M}(r-4 M)+\frac{3 \left(\sqrt{2}-1\right)^{\frac{4-3 \gamma }{\gamma -1}} \left[\left(7 \sqrt{2}-10\right) \gamma -7 \sqrt{2}+8\right] }{128 (\gamma -1) M^2}(r-4 M)^2+\cdots\,.
    \end{aligned}
\end{equation}
It is evident that the divergent terms, which are solely present in terms of the 0-order expansion, may be eliminated through the form of differences. Meanwhile, the higher-order terms do not exhibit divergence. Thus, analytically, there is no problem with our formalism of metric as given in Eqs.~\eqref{gsol}, \eqref{fsol} and Appendixes \ref{app:gaugeg}, \ref{app:gaugef}.

\bibliography{reference, references_qian}

\end{document}